\documentclass[12pt,preprint]{aastex}

\newcommand{\tom}{\tilde{\omega}}
\newcommand{\toml}{\tilde{\omega}_{\rm L}}
\newcommand{\rg}{r_{\rm g}}
\newcommand{\rs}{r_{\rm S}}
\newcommand{\ra}{r_{\rm A}}
\newcommand{\bp}{B_{\rm P}}
\newcommand{\bps}{B_{\rm p\star}}

\newcommand{\up}{u_{\rm P}}
\newcommand{\cm}{\rm cm}
\newcommand{\msun}{{\rm M}_{\sun}}
\newcommand{\omf}{\Omega_{\rm F}}
\newcommand{\rl}{R_{\rm L}}
\newcommand{\sigm}{\sigma}
\newcommand{\ma}{M_{\rm A}}

\shorttitle{Magnetic jets in GRBs}
\shortauthors{Ch.~Fendt \& R.~Ouyed}

\begin{document}

\title{Ultra-Relativistic Magneto-Hydro-Dynamic Jets\\
       in the context of  Gamma Ray Bursts}

\author{Christian Fendt\altaffilmark{1}}
\affil{Institut f\"ur Physik, Universit\"at Potsdam, Am Neuen Palais 10,
       D-14469 Potsdam, Germany}
\email{cfendt@aip.de}

\and

\author{Rachid Ouyed}
\affil{Department of Physics and Astronomy, University of Calgary,
       2500 University Drive NW, Calgary, Alberta, T2N 1N4 Canada}
\email{ouyed@phas.ucalgary.ca}

\altaffiltext{1}{Also: 
 Astrophysikalisches Institut Potsdam, An der Sternwarte 16,
 D-14482 Potsdam, Germany}

\begin{abstract}
We present a detailed numerical study of the dynamics and evolution of
ultrarelativistic magnetohydrodynamic jets in the black hole-disk system under
extreme magnetization conditions.
We find that Lorentz factors of up to $3000$ are achieved and derived a modified
Michel scaling ($\Gamma\sim \sigma$) which allows for a wide variation in the
flow Lorentz factor. Pending contamination induced by mass-entrainment,
the linear Michel scaling links modulations in the ultrarelativistic
wind to variations in mass accretion in the disk for a given magnetization.
The jet is asymptotically dominated by the toroidal magnetic field allowing for
efficient collimation.
We discuss our solutions (jets) in the context of Gamma ray bursts and describe
the relevant features such as the high variability in the Lorentz factor
and how high collimation angles ($\sim 0^o-5^o$), or cylindrical jets,
can be achieved. We isolate a jet instability mechanism  we refer to as the
``bottle-neck" instability which essentially relies on a high magnetization and
a recollimation of the magnetic flux surfaces. The instability occurs at large radii
where any dissipation of the magnetic energy into radiation would in principle
result in an optically thin emission.
\end{abstract}


\section{Introduction}
It is widely accepted that the most conventional interpretation of the observed 
GRBs result from the conversion of the kinetic energy of ultra-relativistic
particles (wind) to radiation in an optically thin region. 
The particles being accelerated by a fireball mechanism taking place near the
central engine \citep{good86, shem90, pacz90}.
The prompt $\gamma$-ray emission is probably  induced by
internal shocks within the wind while the afterglow results
from the wind external shock interacting with the surrounding medium.
The Lorentz factor $\Gamma$ of the relativistic wind must reach
high values ($\Gamma\sim 10^2-10^3$) both to produce $\gamma$-rays
and to avoid photon-photon annihilation along the line of sight,
whose signature is not observed in the spectra of GRBs \citep{good86}.
The crucial point is that for a range of plausible parameters, the prompt
occurs above the Compton photosphere \citep{meza02}.

Whether produced locally or originating from the source, large magnetic fields are
required to account for the synchrotron and/or inverse Compton emission from  a
non-thermal population of accelerated electrons produced behind the (internal and
external) shock waves.
In the case of the afterglow emission, the magnetic field has to be locally
generated by microscopic processes \citep{meza93, wije97,thom00}.
In the case of the prompt $\gamma$-ray emission, such a locally generated magnetic
field is also usually invoked \cite{rees94, papa96, sari97}.
Although a large scale field originating from the hidden source could play the same
role \citep{meza93, meza97,tava96}.

Support for a high collimation of these winds is derived from the break in the
light curves of afterglows for long duration gamma ray bursts (e.g. \citet{stan99}).
The brightness of the optical transient associated to GRB990123 showed a break 
\citep{kulk99},
and a steepening from a power law in time proportional to $t^{-1.2}$, ultimately
approaching a slope $t^{-2.5}$ \citep{cast99}. 
The achromatic steepening of the optical light curve and early radio flux decay of 
GRB 990510 are inconsistent with simple spherical expansion, and well fit by jet
evolution.

The high Lorentz factors invoked, the reference to a magnetic field and the high
collimation are ingredients suggestive of magnetized jets at play in these phenomena.
The extreme magnetic fields  led to models using a highly magnetized millisecond 
pulsar. 
However, due to the baryon loading, GRB jets are different from the essentially
baryon-free pulsar wind case (e.g. \citet{usov94}).
Magnetohydrodynamic (MHD) jets emanating from the surface of an accretion disk
surrounding a central object are naturally loaded with baryons and offer a more
interesting case for GRBs.
Necessary currents are maintained given the standard baryon/mass loading which
guarantees the validity of the MHD approximation out to large distances (up to
$10^{20}$ cm; \citet{spru01} and Appendix A in this paper).
The MHD approximation simplifies the mathematical treatment of the problem but on
the other hand internal magnetic energy dissipation mechanisms are to be sought
that would lead to the proper (GRB-like) emission.
The MHD approximation breaks down when the flow reaches the critical radius beyond 
which current can no longer be sustained; we will refer to this radius as the
{\it MHD radius}.
Note that for the pulsar wind case, the MHD approximation breaks down much earlier,
and plasma theories of large amplitude electromagnetic waves are applied to explain
the prompt emission \citep{lyut01}.
Here, we focus on MHD jets emanating from the disk-black hole system
\citep{thom94, meza97}.
We are interested in solutions that account for the crucial features inherent to
GRB models such as
i) the extreme and variable Lorentz factors as in the internal shock model,
ii) the high collimation ($\theta\sim 1^o-10^o$),
iii) the dissipation mechanisms.
We start in Sec.~2 by introducing the reader to the basic concepts of MHD jets
before specifically focusing on the acceleration (Sec.~3) and collimation 
(Sec.~4) mechanisms.
In Sec.~5, we apply our results to GRBs and discuss ultra-relativistic jets within
the standard internal-external shock model of GRBs. 
In Sec.~6, we isolate a mechanism for jet instability occurring at large radii
(beyond the photosphere) for the case of extreme Lorentz factor and associate it
with possible dissipation mechanisms. 
We summarize our results in Sec.~7.

\section{Basic concepts of relativistic MHD jets}
\label{sec2}

\subsection{Magnetohydrodynamic jets}
\label{sec2.1}

The scenario of MHD jet formation \citep{blan82, pudr83, came86} can be summarized
as follows.
The jet is initiated as a slow wind from the inner disk by a process which is not
yet completely understood, in particular its time-dependent character.
Most probably, some disk instability is responsible for ejecting the matter in
the direction perpendicular to the disk surface.
The disk wind is first launched magneto-centrifugally and further accelerated and
(self-) collimated into a narrow beam by Lorentz forces.

Two key parameters which determine the dynamics of relativistic MHD jet flows,
(i) the plasma magnetization $\sigma$ and
(ii) the light cylinder of the magnetosphere $\rl$.
These will be discussed below.

Considerations of stationary MHD flows have revealed that relativistic jets
must be strongly magnetized \citep{mich69, came86, li92, fend96}.
In that case, the available magnetic energy can be transfered over a small
amount of mass with high kinetic energy.
On the other hand, a very strong magnetization may be in conflict with the
MHD assumption lacking a sufficient large amount of electric charges
which are needed to drive the electric current system (see \S 3.6).

Theoretical modelling of jet formation requires to solve the governing
MHD equations.
However, due to the complexity of MHD and the astrophysical boundary conditions
indicated, a completely self-consistent MHD solution for the jet formation
process being compatible with all the generic features
(MHD self-collimation, accretion-ejection mechanism,
magnetic field generation, spatial and time scales  etc.)
does not yet exist.

In this paper, we will concentrate on solutions to the
{\em stationary, axisymmetric, ideal MHD}
equations in the relativistic limit.
Then the jet magnetic field distribution can be described by
a bunch of nested axisymmetric surfaces,
measuring the magnetic flux through a circular area around the symmetry
axis.
$ \Psi = (1/2\pi) \int {\vec{B}}_{\rm P} \cdot d{\vec{A}}$.
The stationary approach has the advantage to obtain a {\em global}
solution for the relativistic jet MHD structure
(e.g. \citet{li93, fend97}) on spatial scales and
with a resolution which cannot (yet) be reached by time-dependent
simulations \citep{kudo98, koid00}.
Long-term (Newtonian) MHD simulations, however, possess the ability
to demonstrate the self-collimation of MHD jets \citep{ouye97, ouye03}.

\subsection{The magnetization parameter}
\label{sec2.2}
The essential parameter for MHD jets is the {\em magnetization}
parameter \citep{mich69},
\begin{equation}
\sigm = \frac{\Phi^2\omf^2}{4 \dot{M} c^3}.
\label{eq_sigdef}
\end{equation}
The iso-rotation parameter $\omf(\Psi)$ is frequently interpreted as the
angular velocity of the magnetic field lines.
The function $\Phi = B_p r^2$ is a measure of the magnetic field distribution
(see \citet{li93}), and $\dot{M}\equiv \pi\rho v_p R^2$
is the mass flow rate within the flux surface.
Equation ({\ref{eq_sigdef}) demonstrates that the launch of a highly relativistic
(i.e. highly magnetized) jet essentially requires at least one of three conditions
 --
a rapid rotation, a strong magnetic field and/or a
comparatively low mass load.

In the case of a spherical outflow ($\Phi = const$) with negligible gas
pressure one may derive the Michel scaling  between the asymptotic Lorentz
factor and the flow magnetization,
\begin{equation}
\Gamma_{\infty} =
\sigm^{1/3}
\label{eq_sigmich}
\end{equation}
\citep{mich69}.
{\it Assuming a constant mass flux across a jet} with magnetic 
flux $\Phi_{\rm jet}$ and
mass flow rate $\dot{M}_{\rm jet}$, the Michel scaling gives
$\Gamma_{\infty} = (\omf\,\Phi)^{2/3} (\dot{M}_{\rm jet})^{-1/3}$.
It must be mentioned already that Eq.\,({\ref{eq_sigmich}) relies on further
constraints.
For a general relation $\Gamma_{\infty} (\sigm)$ the influence of
collimation, gravity and gas pressure\footnote{
In the case of a ``hot wind'', the magnetization is not a free parameter
anymore.}
must be considered.

Depending on the exact magnetic field distribution $\Phi(r,z)$,
in a {\em collimating jet} the matter can be substantially accelerated
beyond the fast point magnetosonic point
\citep{bege94, fend96},
as it is moved from infinity to a finite radius of several Alfv\'en radii.
As a result, the power law index in Eq.\,({\ref{eq_sigmich}) can be
different from the Michel-scaling
(see Sec.~\ref{sec3.5}; \citet{fend96, vlah01}).
If the $\Phi(r;\Psi)$ is decreasing outwards, the asymptotic flow is
dominated by the kinetic energy \citep{bege94}.
Still, the speed of the flow at the fast point follows the Michel scaling
Eq.\,({\ref{eq_sigmich}), $ \Gamma_{\rm FM} = \sigm (\Psi)^{1/3}$.

The function $\Phi(r;z(\Psi))=\Phi(r;\Psi)$ describes the {\em opening} of the
magnetic flux surfaces $\Psi(r,z)$ comparable to the action of a ``magnetic nozzle''
\citep{came89, li92}.
A flux function $\Phi$ constant along the field lines as applied in the
Michel scaling corresponds to a constant opening angle of the magnetic field.

\subsection{Relativistic features}
\label{sec2.3}
At the {\em light cylinder} (hereafter l.c.) the velocity of the
magnetic field lines ``rotating'' with angular velocity $\omf(\Psi)$
coincides with the speed of light.
The l.c. is located at the cylindrical radius
\begin{equation}
\rl(\Psi) = c/\omf(\Psi).
\end{equation}
For a differential rotation of the field line foot points as in an accretion disk,

the cylinder deforms into a light {\em surface} (of a priori unknown 
shape)\footnote{Note that general relativistic effects affect the shape of the
light cylinder.
Frame dragging close to a rotating black hole implies a second light surface.}.
The l.c. has to be interpreted as the Alfv\'en surface in the limit
of vanishing matter density (force-free limit).
Outside the l.c. the magnetic field lines ``rotate'' faster than the speed of
light\footnote{As the field line is not a physical object, the laws of physics
are not violated.}.
The existence of a strong toroidal field component allows the matter, being frozen 
into the field, to {\em slide along} the field which guarantees $v<c$ also in
the region $R>\rl$.

{\it The location of the l.c. determines the relativistic character of the
magnetosphere.
If the light cylinder is comparable to the dimensions of the object
investigated, a relativistic treatment of MHD is required}.

Contrary to Newtonian MHD, in the relativistic case {\em electric fields}
cannot be neglected.
The poloidal electric field component is directed perpendicular to
the magnetic flux surface. Its strength scales with the l.c. radius,
$E_{\rm p} = E_{\perp} = (r/\rl) B_{\rm p}$.
As a consequence of $E_{\rm p} \simeq B_{\rm p}$, the effective magnetic
pressure can be lowered by a substantial amount (Begelman \& Li \cite{bege94}).

A further difference between relativistic and Newtonian MHD is the fact that
the poloidal Alfv\'en speed $u_{\rm A}$ becomes complex for $r>\rl$,
$u_{\rm A}^2 \sim \bp^2\,(1 - (r/\rl)^2)
= \bp^2-E_{\perp}^2$\,.
Therefore, Alfv\'en waves cannot propagate beyond the l.c. and only fast
magnetosonic waves are able to exchange information across the jet.

Note that the l.c. arises as a relativistic effect due to the
{\em rapid rotation} of the magnetosphere.
This has to be distinguished from the {\em fast proper motion} of
matter in poloidal direction leading to relativistic effects
affecting the inertial forces.
For relativistic MHD jets, both features are interrelated.
A rapidly propagating jet must originate in a rapidly rotating source.
This interrelation is parameterized by the Michel-scaling.

The l.c. is essentially a {\em special relativistic} feature.
Close to the black hole horizon {\em general relativity} becomes relevant.
The existence of the l.c. as a natural length scale in relativistic MHD is
not consistent with the assumption of a {\em self-similar} jet structure.
The latter holds even more when general relativistic effects are considered
(see Sec.~4).

\subsection{Magnetization of GRB jets}
\label{sec2.4}
In this paper we discuss the possibility that GRBs are generated by
ultra-relativistic MHD jets.
Essentially, two model scenarios for the jet origin may be considered, a jet
launched from an accretion disk or by a highly magnetized neutron star.
As the properties of the central source driving the jet are yet unknown, any
estimate on the jet magnetization -- magnetic field distribution, mass flow
rate and rotation -- must rely on somewhat hypothetical parameters.
In the following, we will assume that the jet is launched from the accretion 
disk around a collapsed object and estimate the jet magnetization by constraints
from disk theory.
We do not consider the origin of the disk magnetic field.
It could be advected within the disk from the ambient medium, or generated by
a disk dynamo.

We constrain the jet magnetization at the foot point radius $r_{\star}$ of the jet
$ \sigm = (\bps^2r_{\star}^4 \omf^2 / c^3 \dot{M}_{\rm jet}), $
from the disk equipartition field strength $B_{\rm eq}^2 \simeq P_{\rm gas}$.
This value limits the toroidal magnetic field component which can be built up by
the disk differential rotation, $B_{\rm T} \lesssim B_{\rm eq}$ and also for the
poloidal magnetic field amplification by a dynamo process, 
$B_{\rm P} \lesssim B_{\rm T}$.
In the case of a self-similar advection dominated disk with accretion rate
$\dot{M}_{\rm acc}$, we obtain
\begin{eqnarray}
\label{eq_b_eq}
B_{\rm eq} \simeq 7.8\times 10^9 &{\rm G}&\,\,
\!\left(\frac{\alpha}{10^{-2}}\right)^{\!\!-1/2}
\!\left(\frac{M}{\msun}\right)^{\!\!-1/2}\\
&.&\!\left(\frac{\dot{M}_{\rm acc}}{\dot{M}_{\rm E}}\right)^{\!\!1/2}
\!\left(\frac{r}{\rs}\right)^{\!\!-5/4}\!\!\!\!\!,\nonumber
\end{eqnarray}
%
%
where
$\dot{M}_{\rm E}=10\,L_{\sun}/c^2=2.2\times10^{-8}(M/\msun)\,\msun {\rm yr^{-1}}$
is the Eddington accretion rate,
$\alpha$ is the disk viscosity parameter, and $\rs$ the Schwarzschild radius
(\citet{nara98}, see also \citet{fend01}).
%
%
An optically thin standard accretion disk with Thomson opacity gives
a similar value.
A number of GRB jet models discussed in the literature consider magnetic
field strengths of up to $10^{15}$G
(e.g. \citet{lyut01}).
Equation (\ref{eq_b_eq}) shows that such a high field strength can never be
expected from a disk magnetic field for sub-Eddington accretion rates.
Hyper-accreting disks around black holes have been discussed by \citet{poph99}.
These models provide accretion rates of 0.01 to 10
$\msun{\rm s}^{-1}$ and consider efficient cooling by neutrino losses.
As a consequence, the equipartition field strength may reach
$10^{14}-10^{15}$G. We note however that $\alpha$ is unknown
for hyper-accreting disks.

The other parameter in the magnetization is the jet mass flow rate.
As this is unknown, we scale the jet mass flow rate in terms of the disk
accretion rate, $\dot{M}_{\rm jet} \simeq 10^{-3}\dot{M}_{\rm acc}$.
An upper limit for the jet magnetization can be derived considering
the marginally stable orbit as jet origin, $r_{\star} = r_{\rm ms}$.
Here, we find the largest disk magnetic field strength,
together with the most rapid rotation,
$\omf \simeq 1.4\times 10^4\,(M/\msun)^{-1}(r_{\star}/3\rs)^{-3/2}$
(in the Schwarzschild case).
For a maximally rotating black hole, $(a/M)\simeq 1$,
we have $r_{\rm ms} \simeq \rg$,
where $\rg$ is the gravitational radius of the black hole.
With the maximum jet magnetic flux constrained by the disk equipartition
field, $B_{\star} \simeq B_{\rm eq}$,
the jet magnetization essentially depends on two parameters --
the mass ejection rate and the jet origin,
\begin{equation}
\sigm \simeq 10^5
\left(\frac{\alpha}{10^{-2}}\right)^{-1}\,
\left(\frac{\dot{M}_{\rm jet}}{10^{-3}\dot{M}_{\rm acc} }\right)^{-1}
\left(\frac{r_{\star}}{3\rg}\right)^{-3/2}\ .
\end{equation}
From the Michel-scaling, Eq.~(\ref{eq_sigmich}), we derive a minimum asymptotic
jet Lorentz factor $\Gamma_{\infty} \simeq 10 $.

GRB afterglow observations indicate a total baryonic mass in the burst
of about $10^{-6}\msun$ \citep{pira99},
implying a hypothetical jet mass flow rate of about
$\dot{M}_{\rm jet} \simeq 10^{-7}\msun\,{\rm s}^{-1}$ if the burst lasts for
10\,s.
Such mass flow rates can never be achieved for disk accretion rates
constrained by the Eddington limit,
\begin{eqnarray}
\label{eq_b_mdot}
\dot{M}_{\rm jet} & \simeq & 1.4\times10^{-17}\,\msun\,{\rm s}^{-1}
\,
\left(\frac{\alpha}{10^{-2}}\right)^{-1}
\left(\frac{M}{\msun}\right)\\
& \,&
\quad \quad \quad 
\cdot
\left(\frac{\dot{M}_{\rm acc}}{\dot{M}_{\rm E}}\right)
\left(\frac{r_{\star}}{\rs}\right)^{-3/2}
\left(\frac{\sigm}{1000}\right)^{-1}. \nonumber
\end{eqnarray}
The hyper-accreting stages of accretion disks discussed by \citet{poph99} are
a way out of this dilemma.
With $\dot{M}_{\rm jet} \simeq 10^{-3}\dot{M}_{\rm acc}$,
the inferred accretion rate could be $\lesssim 10^{-2}\msun\,{\rm s}^{-1}$
(necessary for neutrino cooling)
and the equipartition field strength is increased substantially.
However, the flow magnetization governing the asymptotic speed
will remain about the same.

\section{Acceleration -- the asymptotic Lorentz factor}
\label{sec3}
Assuming axisymmetry and stationarity, the equations of ideal MHD can
be re-written into two equations describing the force-balance along
the field (the {\em MHD wind equation}, hereafter WE)
and the force-balance across the field
(the {\em Grad-Shafranov equation}, hereafter GSE).
In general, both equations are interrelated as the source term of the
GSE depends on the dynamics of the MHD wind solution.
In turn, the wind acceleration depends on the magnetic field structure
which is given by the solution of the GSE.
However, in the case of highly relativistic (i.e. highly magnetized)
jets, the influence of the moving matter on the magnetic field can be
neglected
and the field structure may be calculated by the {\em force-free}
GSE (see Sec.~\ref{sec4}).

In this section we present ultra-relativistic MHD solutions of the
wind equation.
For simplicity,
we will consider the {\em cold} wind equation in Minkowski space-time,
which gives us the freedom to investigate the flow dynamics for a
different choice of magnetization.
For a hot wind MHD solution in Kerr metric we refer to 
\citet{fend01}.

As the kinetic time scale for the GRB jet propagation, $\tau_{\rm kin}$,
is well above the time scale when the jet crosses the collimation region,
$\tau_{\rm coll}$,
\begin{equation}
\label{eq_tau}
\tau_{\rm coll}\simeq \frac{10\rl}{c} \simeq 3\times10^{-3}{\rm s}
<<
\tau_{\rm kin} = \frac{10^{12}\cm}{c} \simeq 33\,{\rm s}
\end{equation}
stationarity may indeed be applied for the jet formation region.

\subsection{The relativistic MHD wind equation}
\label{sec3.1}
Combining the MHD equation of motion with the conservation laws for energy
$E$, angular momentum $L$, magnetization $\sigma$ and iso-rotation $\omf$,
we obtain the {\em wind equation} for the poloidal velocity
$\up = \Gamma v_{\rm p}/c$ in Minkowski space-time,
\begin{eqnarray}
\label{eq_wind}
\lefteqn{u_{\rm P} ^2 + 1  = }  \\
& & E^2\,\frac {x^2(1-M^2-x_{\rm A}^2)^2
-(x^2(1-x_{\rm A}^2)-x_{\rm A}^2M^2)^2}{x^2(1-M^2-x^2)^2},\nonumber
\end{eqnarray}
\citep{came86}.
Here, $x_{\rm A} ^2 =(\omf\,L / E)$ defines the Alfv\'en radius $x_{\rm A}$
and $\ma^2 = (4\pi \mu n' u_p^2)/B_p^2$ the Alfv\'en Mach number
$M_{\rm A}$ ($n'$ is the proper particle density).
In the cold wind limit the wind equation simplifies to a polynomial
equation of degree of four for the poloidal velocity.
The polynomial coefficients explicitly depend on the magnetization $\sigm $,
the flux tube function $\Phi $, and the flow parameters energy $E$ and
angular momentum $L$ (see \citet{fend96, fend01}).

At the magnetosonic points the wind equation becomes singular.
A finite solution only exists if numerator and denominator vanishes
together.
For this {\em critical} wind solution the poloidal velocity of the matter
equals the speed of the magnetosonic waves at the magnetosonic points.
We consider such a solution as a {\em global} solution as it is
accelerating from low velocities at small radii to large speed at large
radii.

The cold wind solution is defined by the following parameter set.
The magnetization $\sigm$ and the iso-rotation $\omf$ are free parameters
and may be constraint by the astrophysical boundary conditions.
The total energy density of the flow $E$ is constrained by the regularity
condition at the fast point.
With $\omf$ and $E$ also the total angular momentum flow $L= E/\omf$
is determined.

\subsection{The ultra-relativistic asymptotic MHD jet}
\label{3.2}
Figure \ref{fig_mhd_jet} shows a sample of MHD jet solutions
for a parameter set as motivated above\footnote{
The figure shows the two positive solution branches out of the set of
four numerical solutions of the wind equation at each radial point.
The branch which starts with low velocity at small radius and accelerates to
high speed at large radii is the continuous branch of the (stationary)
{\em physical wind solution}.
The other one decelerates with radius and is not defined for each radius.
For an appropriate parameter set, the latter branch may turn into a
continuous {\em accretion branch}.
The intersections of these branches identify the magnetosonic points
(the Alfv\'en point at $R\simeq\rl$ and fast magnetosonic point at $R>\rl$)
which determine the solution.
}.
We have calculated the flow dynamics for different magnetization
$\sigm = 1000$ and $\sigm = 5000$, and for a different
magnetic field distribution
$\Phi(r;\Psi) \sim r^{-q}$ with $q = 0.01, 0.1, 0.2$.
The light cylinder is located at $\rl = 10^7$cm.
Gravity is unimportant in the cold limit.
Thus, the degree of collimation does not change the character of the
solution as a function of radius $r$ (e.g. $\up(r)$).
However, for a collimated flow, for each radius $r$ the distance from
the source $z(r)$ is increasing with increasing degree of collimation.

Our solutions demonstrate that ultra-relativistic velocities can
be achieved by a MHD jet if it is highly magnetized.
We obtain velocities up to $\up \simeq \Gamma \simeq 3000$ for $\sigm = 5000$.
From the sequence of plots in Fig.\ref{fig_mhd_jet} it can be seen that
magnetization $\sigm$ and field distribution (parameter $q$) play an equally
important role concerning the jet acceleration.
A similar gain in asymptotic velocity can be achieved by either
increasing the magnetization by a factor of five {\em or} increasing
$q$ from close to the Michel value to $0.1$ (compare upper right with
middle left figure).
This confirms earlier results by \citet{bege94} and \citet{fend96}.

\subsection{Pitch angle of the magnetic field}
\label{sec3.3}
It is well known from MHD wind theory that the motion of matter along poloidal
magnetic field lines also implies induction of a strong {\em toroidal} field
component -- so strong that it overcomes the poloidal component for radii larger
then the Alfv\'en radius.
This has been shown early in the case of non-relativistic self-similar MHD
disk winds \citep{blan82},
but holds also for jet flows launched from the disk around a rotating black
hole \citep{fend01}.
For the MHD jet solutions presented in the present paper we find the same
result (not shown).
In particular, the latter paper shows a power law distribution for the toroidal
magnetic field decay in radial direction for radii larger than the light cylinder,
$B_{\rm T} \sim r^{-1}$.
The poloidal magnetic field strength, however, decays faster, as $q \lesssim 0$ for
any reasonable field distribution.
For a monopole type field we have $q = 0 $ and $B_p\sim r^{-2}$ while for a
dipolar field $B_p \sim r^{-3}$ and  $q = -1$.
For the solutions presented here, for the ``asymptotic'' domain
(i.e. for $r=10^4, z\simeq10^7$) the numerical solution gives
$B_{\rm T}/B_p \simeq 10^4$.

{\it Essentially, this proves again the well known characteristics of a
magnetohydrodynamic jet which is asymptotically dominated
by the {\em toroidal magnetic field} component.
This may have important implications
e.g. for the magnetic field distribution in the asymptotic shocks
and for the interpretation of the polarization structure in the
afterglow observations} (see Sec.~\ref{sec6}).

\subsection{Energy balance along the flow}
\label{sec3.4}
MHD jets essentially live from the exchange of magnetic and kinetic energy.
When the jet is launched with low velocity, the energy content is mainly
in the magnetic part, i.e. we have a {\em Poynting dominated} flow.
As the flow accelerates, it gains kinetic energy converting Poynting flux
into kinetic energy flux by Lorentz forces.

Figure \ref{fig_ener} shows the energy partitioning of our MHD jet solutions
for high magnetization $\sigm = 5000$. 
We see that only in the case of $q\simeq0$ the flow remains Poynting dominated
also for large radii.
In the case of a faster magnetic flux divergence ($q>0$), the flow
accelerates substantially beyond the fast surface, converting more and more
Poynting flux into kinetic energy.
For the chosen magnetization the kinetic energy becomes substantial only 
beyond the fast magnetosonic point.
Eventually, the energy distribution in the asymptotic flow is in rough 
equipartition between the kinetic and magnetic contributions.
This is consistent with the claim of \citet{bege94}, but also with (Newtonian)
numerical simulations of jet formation \citep{ouye97, fend02, ouye03}.

\subsection{Asymptotic velocity -- a modified Michel scaling}
\label{sec3.5}
Another feature demonstrating the influence of a divergent magnetic field is the
correlation  (Michel scaling)
between asymptotic jet velocity and magnetization (Fig.~\ref{fig_sig}).

As a measure for the asymptotic ($z \rightarrow \infty$) poloidal velocity
$u_{p,\infty}$, we have taken the velocity at the $r = 10^4\,\rl$.
%
The actual asymptotic value might be a factor of two larger,
however, what is important is that the velocity profile saturates
significantly beyond 10 -- 100 l.c. radii.
This implies that, as far as the efficiency of magnetic acceleration is concerned,
we do not have to consider the far distant region of the flow.

Figure \ref{fig_sig} shows three curves.
The lowest curve corresponds to the original Michel scaling
$u_{\rm p,\infty} \sim \sigm^{1/3}$ in the case of a magnetic field
distribution with $q\simeq 0$.
If the magnetic flux decreasing faster ($q > 0$),
there is a substantial gain in asymptotic velocity.
For the solutions with the decreasing magnetic flux (two upper curves in
Fig.~\ref{fig_sig}) we find a {\em modified Michel scaling},
comparable to an almost linear relation $u_{\rm p,\infty} \simeq A \sigm$,
where $A$ depends on the choice of $q$ (this is crucial since large variations
of the jet's Lorentz factor are possible in this solution; see \ref{sec6.1}).
For the model parameters discussed here, we find $A \simeq 10^{-1/3}$ for 
$q=0.1$ and $A \simeq 10^{-1/5}$ for $q=0.2$.

Note that the interrelation $u_{\rm p,\infty} (\sigm )$ shown in
Fig.~\ref{fig_sig} essentially provides a link between an
asymptotic, ``observable'' jet parameter (velocity) and a
parameter which is intrinsic to the jet origin (magnetization).

\subsection{Applicability of the MHD approximation}
\label{sec3.6}
With the advantage of our approach of knowing the exact solution of the MHD
equations along the collimating jet flow, we are able to check self-consistently
{\it a posteriori} whether the {\em approximation of magnetohydrodynamics} is
satisfied within the calculated flow.
The problem is hidden in the fact that for a cold MHD jet one may find
arbitrarily high velocities for an arbitrarily high flow magnetization.
However, an arbitrarily high magnetization may be in conflict with the intrinsic
{\em MHD condition} under which the solution has been calculated and which
requires a sufficient density of charged particles in order to be able to drive
the electric current system \citep{mich69}.
Below a critical particle density the concept of MHD breaks down.
As a good measure for this critical density we consider the Goldreich-Julian
density $n_{\rm G}$ \citep{gold69, lyut01}.

Following the notation of equation (\ref{eq_wind}), the particle density
$n$ in terms of the Goldreich-Julian density $n_{\rm G}$ as a function of
radius along the magnetic field line is
\begin{eqnarray}
\label{eq_n_ngj}
\frac{n(r)}{n_{\rm G}(r)} & = & \frac{10^7}{\sigma^2} M_A^{-2}(r)
\left(\frac{B_z(r)}{B_z(r_{\star})}\right)^{-1}  \\
& \,& \cdot
\left(\frac{r_{\star}}{0.05\rl}\right)^4
\left(\frac{B_{\star}}{10^{12}{\rm G}}\right)
\left(\frac{R_{\rm L}}{10^{7}{\rm cm}}\right) \nonumber
\end{eqnarray}
(see Appendix).
The index $\star$ denotes a number value at the foot point radius of the
magnetic field line $r_{\star}$.
The magnetic field $z$-component decreases along the opening magnetic
flux surfaces
while the Alfv\'en Mach number increases as the jet accelerates.
In the Appendix (Fig.~\ref{fig_n_ngj}) 
we show some example curves of the relative density
($n(r)/n_{\rm G}(r)$) derived for the jet solutions shown in
Fig.~\ref{fig_mhd_jet}.
Figure \ref{fig_n_ngj} demonstrates that the break-down of the MHD concept
is critically important rather for highly magnetized jets with weak absolute
Poynting flux (implying a low mass flow rate).
For all solutions presented in this paper $n/n_{\rm G}$ stays larger than
about 1000.

\section{Collimation and the compactness problem}
\label{sec4}
\noindent
In this section we discuss the structure of the collimating GRB jet.
We apply a stationary approach and present axisymmetric solutions of the
MHD equations in Kerr metric.
Time-dependent general relativistic MHD simulations of jet formation in the
literature have failed so far to span time periods of more than some
rotational periods.

The axisymmetric magnetic field structure of a stationary collimating MHD jet
follows from the solution of the Grad-Shafranov equation (see above).
The problem is to obtain a {\em global} solution which, at the same time,
also considers the {\em local} force-balance.
So far, fully self-consistent solutions of the stationary relativistic MHD
equations have not yet been able to obtain.
The assumption of a self-similar MHD jet is in general a powerful approach in 
order to obtain self-consistent MHD jet solutions\footnote{
Still, one has to keep in mind that self-similarity implies further constraints
to the solution.
(i) It does not allow to include the jet axis in the treatment.
(ii) Self-similar jets have an infinite radius.
As noted already by \citet{blan82} the radially self-similar assumption becomes
(iii) ``increasingly artificial'' when the jet has formed at large distances 
from the disk}
(see \citet{vlah03} for an application to GRB).
However, for the case of relativistic jets this assumption implies a certain
angular velocity at the foot point of the field lines $\omf(\Psi)\sim r^{-1}$
\citep{li92}
which is in clear contradiction both with the Keplerian rotation of a disk or
with the rigid rotation of a central body.
Similarly, also the magnetization $\sigma$ must be flux-independent.
It is therefore essential to treat the relativistic MHD jet in a non
self-similar, fully two-dimensional approach.
So far this has been possible only in the limit of {\em force-free}
force-balance, neglecting the inertial back-reaction of the matter on the
field structure
\citep{came87, fend97, ghos00}.
It is clear that the previous made comments on a self-similar jet structure become
even more valid in the case of a {\em general relativistic} treatment of MHD jets.
Resulting solutions for ``self-similar relativistic MHD jets''
\citep{cont94, vlah01, vlah03} will not be free of these problems.

\subsection{Structure of relativistic MHD jets from rotating black holes}
Here we summarize the essential steps in calculating the axisymmetric
force-balance of relativistic MHD jets
(for a detailed discussion see \citet{fend97}).
In difference from the previous section we now consider the governing MHD
equations in the framework of {\em general relativity}.

We apply Boyer-Lindquist coordinates in the 3+1 split of space-time
around a rotating black hole of mass $M$ and angular momentum per unit
mass, $a=J/Mc$ with the line element
\begin{equation}
ds^2 = \alpha^2c^2dt^2 - \tom^2\,(d\phi -\omega dt)^2
-(\rho^2/\Delta)\,dr^2 - \rho^2\,d\theta^2.
\end{equation}
$t$ denotes a global time in which the system is stationary, $\phi$ is
the angle around the axis of symmetry, and $r,\theta$ are similar to
their flat space counterpart spherical coordinates\footnote{
The parameters of the metric tensor are defined as usual,
$ \rho^2\!\equiv r^2 + a^2 \cos^2\theta, \,
\Delta\!\equiv r^2 - 2 G M r /c^2 + a^2,$ \\
$\omega\!\equiv  2 a G M r / c\,\Sigma^2, 
 \Sigma^2\! \equiv (r^2 + a^2)^2 -a^2\Delta\sin^2\theta,\,$\\
$\tom \! \equiv (\Sigma/\rho)\,\sin\theta,           \,
\alpha \! \equiv  \rho\,{\sqrt{\Delta}} / \Sigma $
}
Here, $\omega $ is the ``frame dragging'' angular velocity of an observer
with zero angular momentum (ZAMO), $\omega = (d\phi/dt)_{\rm ZAMO}$.
The lapse function $\alpha $ describes the lapse of the proper time $\tau$
in the ZAMO system to the global time $t$, $\alpha = (d\tau/dt)_{\rm ZAMO}$.

We define the axisymmetric magnetic flux $\Psi$ through a loop of the
Killing vector $\vec{m} = \tom^2\nabla\phi$ as
\begin{equation}
\label{eq_psi_k}
\Psi (r,\theta) = \frac {1}{2 \pi} \int {\vec {B}}_{\rm P} \cdot d{\vec{A}}\,,
\quad\quad {\vec{B}}_{\rm P} = \frac{1}{\tom^2}\nabla \Psi \times {\vec{m}},
\end{equation}
(see Sec.~\ref{sec2.1}).
The indices P and T denote the poloidal and toroidal components of a vector.
Similar to Eq.\,(\ref{eq_psi_k}) the total poloidal electric current is
defined by
$ I = -\int \alpha {\vec {j}}_{\rm P} \cdot d{\vec{A}}
  = - \frac{c}{2}\alpha\tom B_{\rm T}\,. $
In a force-free magnetosphere with
$ 0 = \rho_{\rm c} \vec{E} + \frac {1}{c} \vec{j}\times\vec{B}$
the poloidal electric current is parallel to the
poloidal magnetic field $\vec{B}_{\rm P} \parallel \vec{j}_{\rm P}$
and is a conserved quantity along the magnetic flux surfaces, $I = I(\Psi)$.

The axisymmetric force-balance perpendicular to the magnetic field is
described by the Grad-Shafranov equation (hereafter GSE).
In the limit of highly relativistic (i.e. highly magnetized) jets,
inertial forces have a negligible influence on the {\em structure} of
the magnetosphere and we may apply the force-free limit of the GSE,
\begin{equation}
\label{eq_gse}
\tom \nabla \cdot
\left({\alpha\,\frac{1-\left(\tom 
/\tom_{\rm L}\right)^2}{\tom^2}}\nabla\Psi\right)
= -\frac{g_{\rm I}}{2} \frac{1}{\alpha\tom}\;\frac{dI(\Psi)}{d\Psi}\,.
\end{equation}
For simplicity, differential rotation of the jet basis has been neglected,
$\omf(\Psi) = const. = \omf$.
The two light surfaces are located at the radial position
$\toml = \left(\pm {\alpha}/({\omf - \omega})\right)^{1/2}$.
The $+$ sign holds for the outer light surface with $\omf > \omega$,
while the $-$ sign stands for the inner light surface, where $\omf < \omega$.
The asymptotic radius of the outer light surface 
($\toml$ for $z\rightarrow \infty $),
the light cylinder, is denoted by $R_{\rm L}$.
Normalizing the GSE (\ref{eq_gse}) using $\tom \rightarrow R_{\rm L} \,\tom$;
$\nabla \rightarrow (1/R_{\rm L}) \,\nabla\,$;
$\Psi \rightarrow {\Psi }_{\rm max}\,\Psi \,$ and
$I \rightarrow I_{\rm max}\,I$
is numerically advantageous, but also provides insight in the physical
characteristics of the solution.
The coupling constant $g_{\rm I}$ measures the strength of the source term
of the GSE,
\begin{equation}
\label{eq_g_i}
g_{\rm I} = \frac {4 I_{\rm max}^2 R_{\rm L}^2}{c^2 {\Psi }_{\rm max}^2} =
0.5\,(\frac {I_{\rm max}} {10^{15} {\rm A}})^2
(\frac {R_{\rm L}}{10^{7}{\rm cm}})^2
(\frac {{\Psi }_{\rm max}}{10^{21}{\rm G cm^2}})^{-2}\!\!.
\end{equation}
Basically, $g_{\rm I}$ determines the strength of the electric current
$I_{\rm max}$.
A high coupling constant corresponds to a strong poloidal electric current
(respectively a strong toroidal magnetic field) and therefore implies a
strong jet collimation.
Note that the coupling between the source term and the poloidal field structure
considers {\em only electromagnetic} quantities.
The solution is, however, calculated in Kerr metric, and considers also
gravity.
The link between the two governing length scales defined in the problem --
the asymptotic light cylinder and the gravitational radius -- is made by
by choosing $\omf$ in terms of the black hole rotation
$\Omega_{\rm H}$ (see below).

Equation (\ref{eq_g_i}) applies parameter estimates as discussed above for GRB
jets launched by black hole accretion disks.
The maximum poloidal electric current in the disk-jet system is given by the
disk equipartition field strength.
For an advection dominated disk (see Eq.~(\ref{eq_b_eq})) we have
$B_{\rm T}^{\rm max} \lesssim B_{\rm eq} \simeq 10^9$G and
$ I_{\rm max}  \lesssim 3\,\rg B_{\rm T} c/2 \simeq 10^{15} {\rm A}.$
Since $B_{\rm P}^{\rm max} \lesssim  B_{\rm T} \simeq B_{\rm eq}$ in the disk,
we derive a maximum magnetic flux from the disk
$\Psi_{\rm max} \lesssim \pi\,(3\rg)^2 B_{\rm P}^{\rm max}\simeq 10^{21}{\rm G\,cm^2}$.
The expression for $g_{\rm I}$ can be further simplified considering that
the light cylinder radius in Eq.\,(\ref{eq_g_i}) is governed by the rotation
of the jet foot point $\omf$.
As both  $B_{\rm T}^{\rm max}$ and $B_{\rm P}^{\rm max}$ are limited
by $B_{\rm eq}$ as a function of radius, we finally obtain
\begin{equation}
g_{\rm I} = 0.6\,\left(\frac{B_{\rm T}^{\rm max}}{B_{\rm P}^{\rm max}}\right)^2
\left(\frac {r_{\star}}{6\rg}\right).
\label{eq_g_i_2}
\end{equation}
For the case of hyper-accreting disks discussed above, the limiting equipartition
field strength constrains the poloidal and toroidal field equivalently.
As a consequence, the coupling $g_{\rm I}$ remains the same.

\subsection{Example solution for a fully collimated GRB jet}
Here we present an example solution for the stationary axisymmetric
force-free magnetic field structure of GRB jets calculated from the
Grad-Shafranov equation (\ref{eq_gse}).
The GSE is solved applying the method of finite elements.
This allows to consider the complex geometrical structure of the
black hole - disk - jet system.
Details of the method of solution are discussed elsewhere \citep{fend97}.
We summarize the following characteristics of the solution.
\begin{itemize}
\item[-] The shape of the collimating outer jet boundary is not known
a priori. It is determined by the internal structure of the jet
and is particularly constrained by the regularity condition along the
outer light surface (i.e. a smooth transition of the magnetic field lines).
\item[-]
The ``free function'' $I(\Psi)$ in the GSE source term is chosen from the
analytical solution of an asymptotic jet in perfect collimation.
\item[-]
The interrelation between the two characteristic length scales of the solution,
the asymptotic light cylinder and the gravitational radius
-- hence, the link between electrodynamics and gravity -- is made by
choosing the angular rotation $\omf(\Psi)$.
Here, $\omf = 0.4 \Omega_{\rm H}$ corresponding to the Keplerian
rotation at the marginally stable orbit.
With that $\rl = 10 \rg$ for $a = 0.8$.
\item[-]
Other parameters are the magnetic flux from the black hole in terms of the
total flux, $\Psi_{\rm BH} / \Psi_{\rm max} = 0.2$,
and the profile of the disk magnetic flux distribution.
\end{itemize}
Figure \ref{fig_2D} shows the resulting structure of the collimating
magnetic flux surfaces.
The solution extends from the inner light surface to the asymptotic regime
of a fully collimated jet.
We find a perfect collimation of the asymptotic jet in agreement with
analytical models of \citep{appl93}.
We find a rapid collimation of the outflow into a cylindrical shape within
$50\rg$ distance from the central black hole.
The expansion rate
(asymptotic jet radius in respect to the jet foot point radius)
is about 10.
The asymptotic jet radius is 5 l.c. radii.
Due to the numerical resolution we do not obtain solutions with larger jet 
radius.

Whether the jet is self-collimated by magnetic tension or pressure
collimated by an ambient medium cannot be answered by this approach.

An interesting fact is that the flux surfaces emerging from the
black hole (i.e. from the inner light surface in our approach)
asymptotically partly intersect with the outer light surface.
This implies that the mass load in the jet provided by the accretion disk
is predominantly in the outer layers of the jet at radii $r>\rl$.
Here, we expect the high velocities calculated in the previous section
together with a strong toroidal magnetic field.
The inner part of the jet with flux surfaces emerging from the
black hole, should be dilute of matter, i.e. we have a hollow jet stream.
This part of the jet may be powered by the electromagnetic interaction with
the black hole itself (Blandford-Znajek mechanism).

The strength of the coupling is similar to the case of active galactic
nuclei (AGN).
As the coupling constant Eq.\,(\ref{eq_g_i}) measures the strength of
the collimating toroidal magnetic field strength in terms of the
de-collimating effects poloidal field and rotation\footnote{Note that
the coupling constant corresponds to the inverse magnetization for
force-free jets}.
Therefore, we may expect a similar structure of the jet magnetosphere
in both examples.
A strong coupling implies efficient collimation;
we find that collimation angles as low as
$0^o-5^o$ can be achieved with proper tuning
of parameters.
From AGN jets we know that they are well collimated with asymptotic jet
radii up to of about $100\,\rl$.
Just from the similarity of the coupling constant we might expect the
same scaling also for jets in GRB's.
To calculate such large jet radii is, however, beyond the capability of
our numerical code.

\section{The ``bottle-neck'' instability for highly magnetized jets}
\label{sec44}
In the previous sections we have presented stationary solutions for
highly magnetized MHD jets -- the flow dynamics along the collimating
magnetic field and the axisymmetric magnetic field structure.
Our time scale estimates (Eq.\,\ref{eq_tau}) indicate that the stationary
approach may indeed be applied when considering the jet formation region.
Nevertheless, it is clear that the GRB itself must be the consequence
of a time-dependent process.
The current picture of GRB invokes highly relativistic shells moving with
different speed, catching up with each other by forming a highly energetic
shock front -- the gamma ray burst \citep{pira99}.

From our MHD solutions we have found indication of a possible excitation
mechanism for a flow instability.
This instability essentially relies on (i) a {\em high magnetization} and
(ii) a {\em re-collimation} of the governing magnetic flux tube.
The solutions discussed in Sec.~3 have been calculated for a
{\em decreasing} function $\Phi$ along the jet corresponding to
a increasing opening of the flux tube.
However, the two-dimensional jet magnetic field structure as presented
in Sec.~4 shows an interesting property.
The global jet collimates into a cylindrical shape with
the single flux surfaces turning from an initially conical outflow
close to the disk into an alignment parallel to the jet axis.
This implies a {\em re-collimation of the single flux tube} -- easily
visible when comparing the width of the flow channel along the
outer flux surfaces (see also \citet{fend96}).
The maximum width is typically located in the region of the strongest
curvature at a distance of several $\rl$ from the central source.

Solving the wind equation by considering a re-collimation of the
flux tube we find that {\em no stationary wind solutions exists beyond
the radius where re-collimation happens}.
We cannot directly answer the question what exactly happens there as
this is beyond our stationary approach.
For comparison, one may think of the earth magnetosphere reflecting solar
wind particles from the polar region or the more general picture of a
``magnetic bottle'' of converging magnetic field lines.

In the appendix we show a typical MHD wind solution as discussed in Sec.~3,
but modified for the re-collimation feature by prescribing an increasing
magnetic flux tube function $\Phi$ beyond a certain radius.
In this example we applied a cosine-like recollimation of the flux tube
$\Phi(r)\sim r^{-0.2}$ within a certain radius range beyond 
$r > 50 \ra \simeq 50 \rl$ (see Eq.~\ref{eq_recol}).
Figure \ref{fig_bottle} shows an example velocity distribution along the
field line for different magnetization.
For high magnetization the magnetic field is strong enough to brake the
matter sufficiently.
The exact dynamical behavior depends on the parameters of the wind solution.
Essentially, the re-collimation leads to a deceleration of the wind flow
before the stationary solution ceases to exist.
Applying a spatially limited region of re-collimation, e.g. a sinusoidal
variation of $\Psi(r)$ between two radii, there may exist a unique
{\em stationary} flow solution which can be decelerated below the
magnetosonic velocity.
However, in reality such a unique combination of field distribution
and mass flow rate may not exist.

Therefore, we propose the above mentioned indication of a
``bottle-neck'' instability as a generic reason for the velocity
variation and the formation of shock waves in GRB.
This instability is working predominantly in the outer layers of the jet
and in the collimation regime at distances of several l.c. radii from the
central object.

The onset of the instability depends on the details of the flow propagation.
However, in general, the essential conditions are {\em strong magnetization}
and {\em re-collimation}.
Due to the Michel scaling, this corresponds in particular to high Lorentz
factors as present in jets of GRBs.
In our example model we compare magnetizations of $\sigma = 10, 1000, 5000$
and a certain sinusoidal variation of the flux tube function (see Appendix).
It can be seen that for low magnetization a stationary solution exists from
the jet foot point to the asymptotic regime.
For higher magnetization, the stationary solution ceases to exists at a
certain radius.
For the parameter choice applied here, we do not find a stationary flow domain
around the location of local maximum re-collimation.
In the case of further recollimation, no stationary wind solution will be found
beyond this radius.
In our case, with an again increasing size of the flux tube, stationary
solution can be found beyond the non-stationary regime.
This, however, has no consequence as the jet flow has become non-stationary 
already. {\it We note that it is not straightforward to estimate
the exact location of the radius where the instability occurs. However,
its seems to occur at large enough radii to assume that it
is beyond the Compton photosphere.}

\section{Ultra-relativistic MHD jets in the GRB picture}
\label{sec6}

\subsection{Varying Lorentz factor and internal shocks}
\label{sec6.1}

\noindent
\underline{Mass-load}

Variations in the Lorentz factor of $10 - 1000$
($10^2 < \Gamma < 10^5$) as suggested by the GRB
internal shock model are hard to obtain from the original Michel scaling.
In fact, $u_{\rm p,\infty} \sim \dot{M}_{\rm jet}^{-1/3}$
would imply a variation in the mass load by several orders of magnitude more.
The existence of a modified Michel scaling as described in \S \ref{sec3.5}
(see Fig.~\ref{fig_sig}) leaves the possibility of having a substantial change
on the Lorentz factor by the variation of the initial flow magnetization.
A variation in the jet magnetization can be caused by a change in the mass
injection rate from the accretion disk into a jet with temporarily constant
magnetic field implying $u_{\rm p,\infty} \sim \dot{M}_{\rm jet}^{-1}$. 
It is interesting to note that these variations are directly induced by
conditions at the source pending contamination by the surrounding material
as discussed next.

\noindent
\underline{Ambient\ mass-entrainment}

One would expect parts of the jet to be contaminated by the ambient material
\citep{daig00} as to cause local variation to the intrinsic Lorentz factors
(as given by the disk mass loading or magnetization).
In fact, it is likely that the variations in the Lorentz factors are induced
by a combination of varying mass-load and mass-entrainment.
This would result in a multiple shock mergers inducing GRBs (as in the internal
shock scenario) which can be seen if mass entrainment occurs mainly beyond 
the compactness radius.

\subsection{Magnetic energy dissipation}
\label{sec6.2}
It is argued that, for models of internal shocks in GRBs to successfully
reproduce the GRB temporal features,
different shells of matter should have a comparable energy and their
different Lorentz factors should arise due to modulation of the accelerated
mass (\citet{pira01} and references therein). 
In MHD jets, the energy (mainly magnetic at the base of the flow) is roughly
constant as the magnetization is not expected to vary much. 
A variable mass-load combined with the entrained mass, as we have said,
offers the modulation needed to account for the wide range in the Lorentz
factors\footnote{This may also lead to a narrow range in the Lorentz factor
distribution as part of the jet with higher Lorentz values would in principle
induce higher entrainment; a notion which remains to be confirmed.}.
By itself, such a variable source is not enough to explain the variable light
curves since magnetic energy dissipation mechanisms will not necessarily be
efficient or adequate in reproducing the GRB temporal
structure\footnote{Dissipation in shocks (internal and external shocks) is
fundamentally different from  magnetic energy dissipation in MHD outflows
where non-thermal electron distributions is not guaranteed.}.
For the Crab Pulsar wind in which energy is transported predominantly as
Poynting flux, the fluctuating component of the magnetic field in such a flow
can in principle be dissipated by magnetic reconnection and used to
accelerate the flow \citep{kirk03}.
In our model, such mechanisms might be at play as a result of the
``bottle-neck'' instability.
A conversion of the fraction of the energy of the accelerated particles into
radiation is interesting within the GRB context.
The pulsation of the emitted radiation in our picture, as we have argued above,
would be linked to the variation in the accretion rate at the source.
We note however that it is not yet clear how much of the magnetically
dissipated energy can be seen in the form of $\gamma$-rays \citep{spru01}.

\subsection{Collimation and polarization}
\label{sec6.3}
The results from \S \ref{sec4} suggest that ultra-relativistic MHD jets are
highly collimated. 
Extrapolating from what we know in the case of AGN jets this implies that
collimation angles as small as $0^o-5^o$ are feasible.
Such an efficiency in collimation is a statement that the jet is asymptotically
dominated by the {\em toroidal magnetic field} component} (\S \ref{sec3.3}).
This may have important implications e.g. for the magnetic field distribution
in the asymptotic shocks and for the interpretation of the polarization
structure in the afterglow observations.

\section{Summary}
\label{summary}
We studied ultra-relativistic MHD jets in the context of GRBs. 
We have presented stationary solutions for the governing MHD
equations considering the axisymmetric structure of the collimating jet and
the acceleration of matter by the magnetic field.
Special and general relativistic effects have been taken into account.

Lorentz factors up to 3000 can be obtained for suffient strong jet 
magnetization.
The advantage of our approach of actually solving the governing
MHD equations is that we can prove {\it a posteriori} the
applicability of MHD itself by comparing the matter density with the
Goldreich-Julian density along the jet flow.
The asymptotic jet magnetic field is dominated by the toroidal
component.
The energy distribution within the asymptotic jet is almost in 
equipartition between magnetic and kinetic energy.

For the structure of the axisymmetric MHD jet we find a rapid 
collimation
to almost perfect collimation within a distance from the black hole
of about half the jet radius (of 5 light cylinder radii in this example).

Among the features that are crucial to standard models of GRBs, are 
(i) the modified Michel scaling ($u_{\rm p,\infty}\sim \dot{M}_{\rm jet}^{-1}$) 
which allows for a plausible variation in the Lorentz factor by variations in the 
source parameters such as the disk magnetization and/or mass-loading; 
(ii) the high degree of collimation ($\theta\sim 0^o-5^o$)  which is within
the range of values derived from breaks of afterglow light curves.
We isolated a jet instability that develops at extreme Lorentz factors. 
This instability which seems to arise at large radii, if it occurs beyond the
compactness radius, would result on an optically thin emission once its
associated magnetic field is dissipated.
We expect different magnetic field configurations to lead to different 
dynamics (e.g. asymptotic profiles, dissipation efficiency).
This, in principle, could reproduce the large diversity and duration range
of GRBs.

Among the open questions are the extreme magnetic fields required, and the
effect of baryon contamination (via mass-entrainment) on the distribution of the
Lorentz factors.
It is also not clear at this stage why and how the engine
(the underlying hole-disk system here) should stop for several dozens
seconds before bursting again as seen in GRBs.

\appendix

\section{Applicability of the MHD concept}
As discussed above highly relativistic jets must be strongly magnetized
(\citet{mich69, came86, fend96}; see Sec.2.4).
Magnetohydrodynamic jets live from magnetic to kinetic energy conversion.
More kinetic energy per fluid element can be delivered for jets with
correspondingly strong magnetic field, or low mass flow density,
respectively.
On the other hand, the {\em MHD concept} itself requires a minimum density
of charged particles in order to allow for electric currents in the flow.
The jet flow mass density naturally decreases along the flow as the jet
originates in a very small region close to the central source and first
expands in almost radial direction before it collimates into a narrow beam.
When the flow density is below the critical density at  the MHD radius
$r_{\rm MHD}$, the MHD assumption breaks down.
With our model approach, we may determine self-consistently the possible
breakdown of the MHD conception
as our MHD wind solution delivers all the dynamical and electromagnetic
properties of the flow.


\citet{mela96} quantify the range of validity for 
{\em ideal} MHD, $\vec{E} + \vec{v}\times\vec{B}=0$,
by inspecting the one-fluid equations for a cold, neutral plasma.
They consider the generalized Ohm's law for fluctuations of time scales $\tau$
on characteristic length scales $\lambda$ in the magnetohydrodynamic quantities.
In the end, there are four constraints for ideal MHD.
First, it is required that $j<<n e  v$ and $\rho_{\rm c} v <<j$
with the charge density $\rho_{\rm c}$.
If the conduction current if dominating the displacement current,
Maxwell's equations imply that for MHD
$(4\pi j\tau /E) \simeq (c\tau/\lambda)^2$.
With the relation
\begin{equation}
\label{eq_mel_mel}
\frac{\Gamma c^2}{\omega_0^2\lambda^2}{\rm max}
\left(1,\frac{\lambda}{\tau v},\frac{\tau v}{\lambda}\right) << 1
\end{equation}
(with $\omega_0 \equiv \sqrt{4\pi n e^2/m_{\rm p}}$)
Ohm's law actually reduces to $\vec{E} + \vec{v}\times\vec{B} \simeq 0$.
In the case of a pulsar wind with $\lambda \simeq \rl$,
$\omega \simeq \Omega_{\star} = \omf$ and $\tau \simeq \omega^{-1} = \rl/c$,
Eq.~(\ref{eq_mel_mel}) takes the form $\Gamma \omega^2 << \omega_0^2$
which is equivalent to the estimate of Michel (1969).
For the Crab pulsar the MHD radius is located at $r_{\rm MHD} \simeq 10^5\rl$
\citep{mela96}.

A similar criterion for the applicability of MHD is a matter density above
the {\em Goldreich-Julian charge density}
\citep{gold69, mich69}.
The Goldreich-Julian particle density along a magnetic flux surface
$\Psi (r,z)$ is
\begin{equation}
\label{eq_n_gj}
n_{\rm G} = -\frac{(\nabla\Psi \cdot \nabla)(r^2\omf)}{4\,\pi\,c\,e\,r^2}
= \frac{B_z(r;\Psi)}{2\pi\,e\,c\,\rl},
\end{equation}
and is not negligible for rapid (relativistically) rotating magnetospheres.
Again, a particle density below the Goldreich-Julian density indicates that
not enough charges are available to carry the electric current.
From Eq.~(\ref{eq_n_gj}) we see that if the particle density profile decreases
faster than the $B_z$ component,
there may exist a critical distance $r_{\rm MHD}$ where the matter particle
becomes lower than the Goldreich-Julian density.

\citet{usov94} has applied this approach in order to model GRBs generated
by millisecond magnetars of $3\times10^{15}$G dipolar field strength,
\begin{equation}
n_{\rm G} = 4\times10^{16}\frac{r_{\star}}{r}
\left(\frac{B_{\star}}{3\times10^{15}{\rm G}}\right)
\left(\frac{\omf}{10^4 {\rm s}^{-1}}\right)^3 {\rm cm}^{-3}.
\end{equation}
In the case of a leptonic plasma wind Usov derives a critical distance
$r_{\rm MHD} \simeq 10^{13}$cm for the given stellar magnetic field strength
and field distribution and the rotation rate ($\omf = 10^4{\rm s}^{-1}$).
It is, however, another question to extrapolate these variables over more
than six orders of magnitude in distance from the star to the region
where the critical radius is located in this model.
A similar approach has been applied in the context of ``Poynting-dominated''
outflows in GRBs \citep{lyut01},
assuming a certain magnetic field and density distribution.
In this model, large-amplitude electromagnetic waves break at the MHD radius
located at about $10^{14}$cm while accelerating particles to ultra-relativistic
speed.

In difference to the previous work, in our paper we calculate the solution
of the relativistic MHD wind equation from the jet basis to the asymptotic regime
providing us with a set of dynamical properties of the flow.
In particular, this allows us  to check self-consistently (but a posteriori)
the consistency of our solutions with the MHD conception.
With that we may also compare different jet geometries -- rapid or weak
collimation --
and may constrain the parameters of the jet mass loading and the magnetic field
distribution.
We derive an expression for the matter particle density from the definition
of the Alfv\'en Mach number, 
\begin{equation}
\label{eq_ma}
M_{\rm A}^2 = \frac{4\pi \mu n' u_p^2}{B_p^2}
\end{equation}
with the proper particle density $n'$, the gas entropy $\mu$ and
the poloidal velocity $u_{\rm p} \equiv \Gamma v_{\rm p} /c$
\citep{came87, fend96}.
Re-writing Eq.~(\ref{eq_ma}) gives the proper particle density along the 
magnetic field line as a function of the Alfv\'en Mach number,
\begin{equation}
n' = \frac{1}{4\pi} \frac{\Psi_D^2}{m_{\rm p} c^2 \sigma^2 \rl^2 M_{\rm A}^2}.
\end{equation}
Relating that to the Goldreich-Julian density Eq.~(\ref{eq_n_gj}),
we find Eq.~(\ref{eq_n_ngj}).
As the Alfv\'en Mach number generally increases faster
than the z-component of the magnetic field decreases along the flow,
the ratio $(n'/n_{\rm GJ}$) will decrease.
In Fig.~\ref{fig_n_ngj} we show
the profile of the density ratio for four of the example wind solutions 
presented in Fig.~\ref{fig_mhd_jet}.
For all four cases the mass flow density stays above the Goldreich-Julian
density.
The MHD jet parameters are the magnetization $\sigma = 1000, 5000$ and 
the foot point magnetic field strength of
$B_{\rm p} = 10^9{\rm G}, 10^{12}{\rm G}$.
In particular, we find that the density ratio approaches a constant value
$n'/n_{\rm GJ} \gtrsim 100$ for large radii.
This reflects the fact of a {\em jet collimation}.
For a further (substantial) expansion of the jet flow for these choice
of initial parameters, the mass density would become under critical.

\section{Recollimation -- the ``bottle neck'' instability}
Here we show example solutions of the MHD wind equation for the case of
a local re-collimation flux tube function $\Phi(r)$.
For this purpose we modified the decreasing magnetic flux tube function
$\Phi(r)$ by simply adding a cosine shape increase within a certain radius
range along the jet.
For the solutions presented in Fig.~\ref{fig_bottle} we have prescribed
\begin{eqnarray}
\label{eq_recol}
\Phi(r) & \sim & r^{-0.2}
\quad {\rm for}\quad r < 50\,\ra \nonumber \\
\Phi(r) & \sim & r^{-0.2}
\left(1+1.25 \left(1-\cos(\frac{r-50}{2000}\pi) \right) \right) \\
 &\,& \quad\quad
\quad {\rm for} \quad 50\,\ra < r < 2050\,\ra \nonumber \\
\Phi(r) & \sim & r^{-0.2}
\left(1+1.25 \left(1-\cos\pi \right) \right)
\quad {\rm for}\quad r > 2050\,\ra \nonumber
\end{eqnarray}
We have computed the wind solution for several different magnetizations
and otherwise equal parameters.
Figure \ref{fig_bottle} shows that for low magnetization ($\sigma = 10$)
the stationary wind solution reaches from the foot point into the asymptotic
domain. For higher magnetization ($\sigma = 1000, 5000$) no stationary
solution can be found along a certain radial range around the point of
(local) maximum re-collimation (see $\Phi(r)$-plot).

\clearpage


\begin{figure*}

\epsscale{1.1}
\plotone{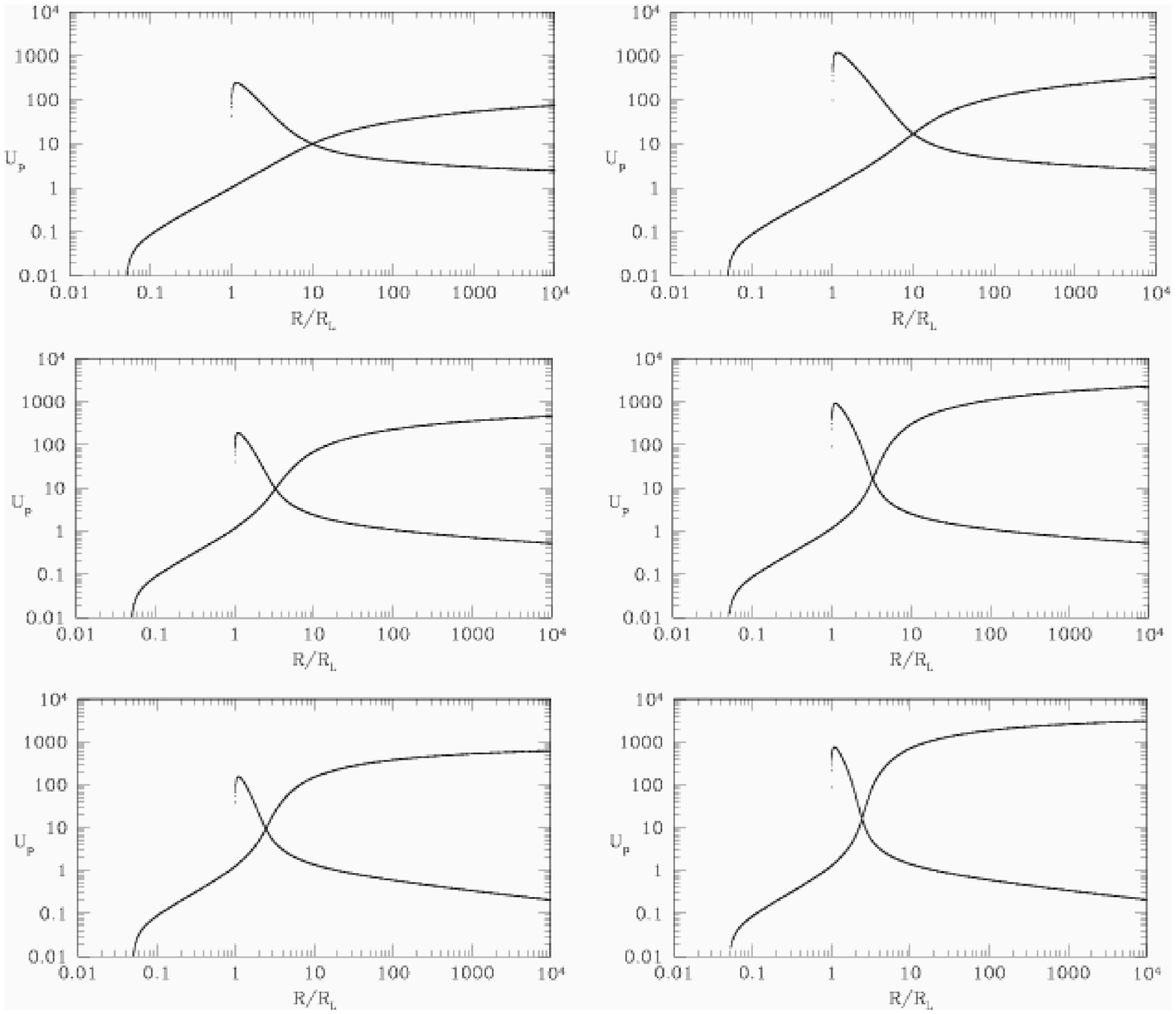}
\caption{Relativistic MHD wind/jet solution. Poloidal velocity $u_{\rm p}$
along the magnetic field line.
Parameter: $\sigm = 1000$ ({\it left}), $\sigm = 5000$ ({\it right}),
Parameter: $\Phi(r) \sim r^{-q}$, $q=0.01$ ({\it top}),
$q=0.1$ ({\it middle}).
$q=0.2$ ({\it bottom}).
\label{fig_mhd_jet}
}
\end{figure*}

\begin{figure}

\epsscale{0.6}
\plotone{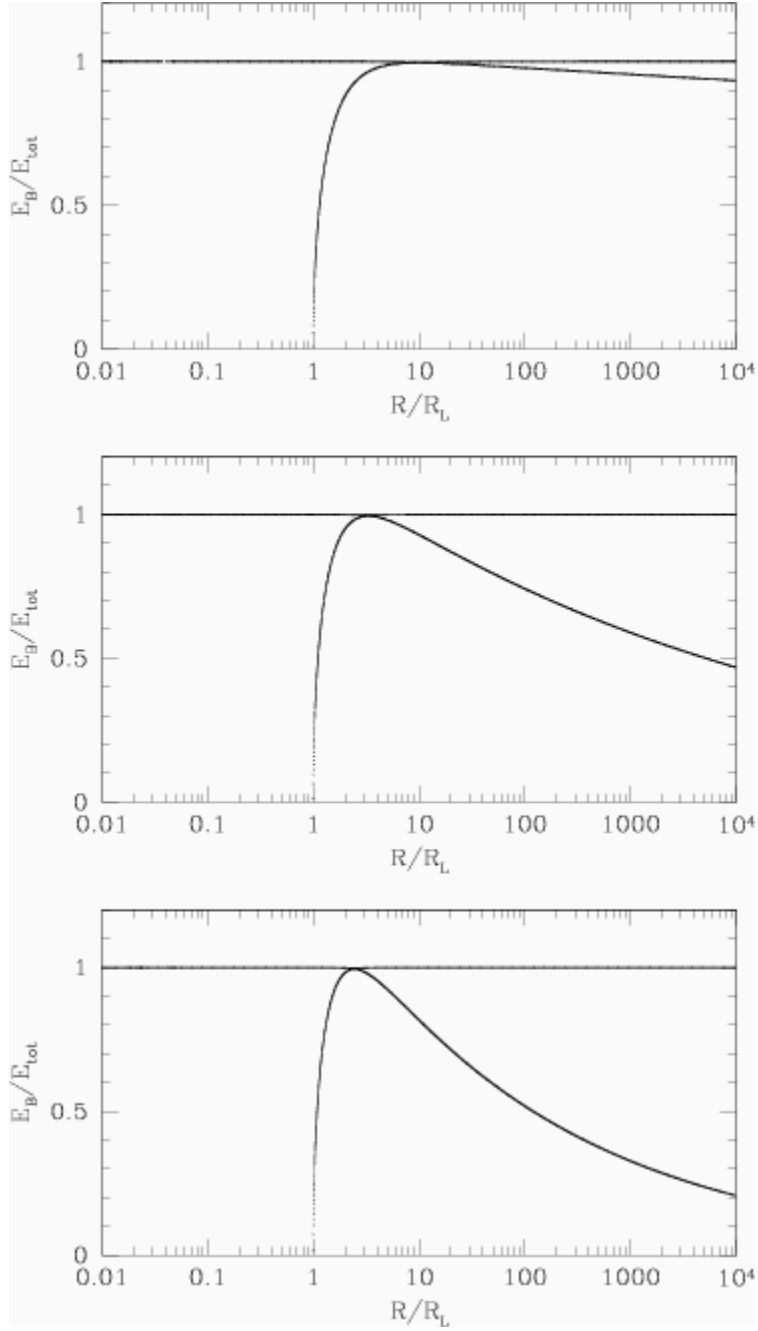}
\caption{Relativistic MHD wind/jet solution. Magnetic energy in terms of
total energy as a function of radius along the magnetic field line.
Note the fast magnetosonic point as intersection of both curves.
Parameter: 
$\sigm = 5000$, 
$\Phi(r) \sim r^{-q}$, $q=0.01, 0.1, 0.2$
(from {\it top} to {\it bottom}).
\label{fig_ener}
}
\end{figure}

\begin{figure}
\epsscale{1.0}
\plotone{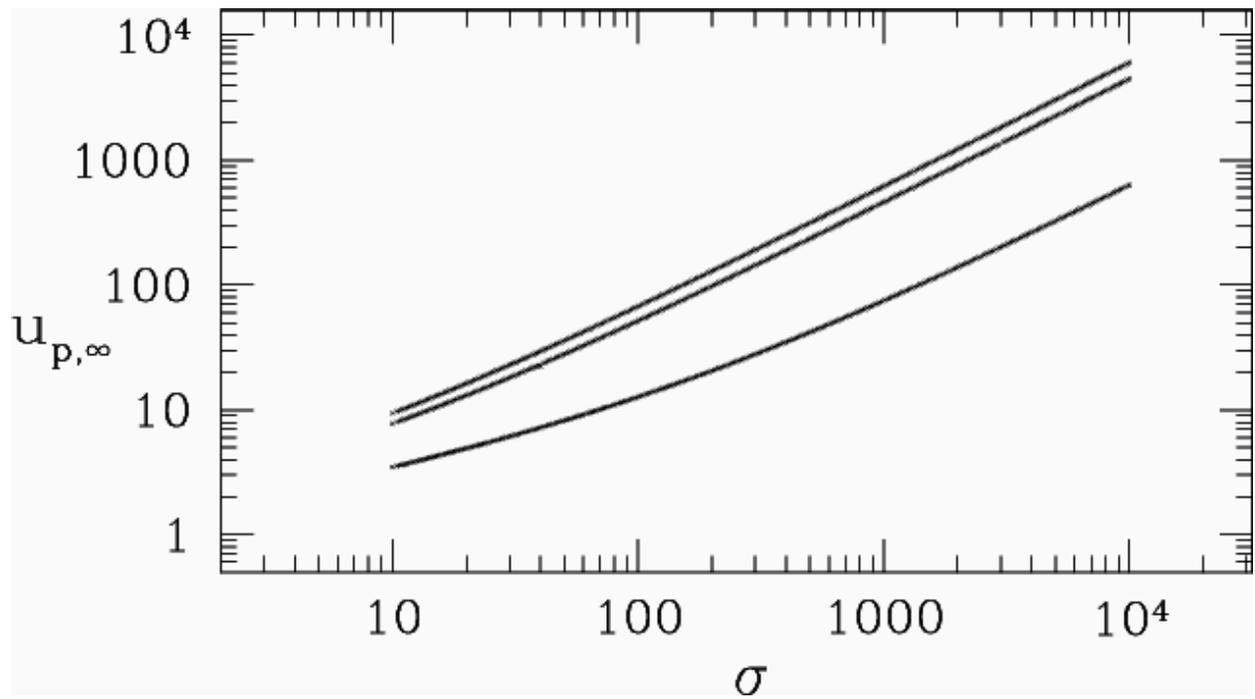}
\caption{
Relativistic MHD jet solution. Modified Michel scaling
$u_{\rm p,\infty} (\sigm)$ for a different choice of the magnetic field
distribution $\Phi(r;\Psi) \sim r^{-q}$ with
$q=0.2$ ({\it top curve}),
$q=0.1$ ({\it middle curve}),
$q=0.01$ ({\it bottom curve}).
\label{fig_sig}
}
\end{figure}

\begin{figure}
\epsscale{1.0}
\plotone{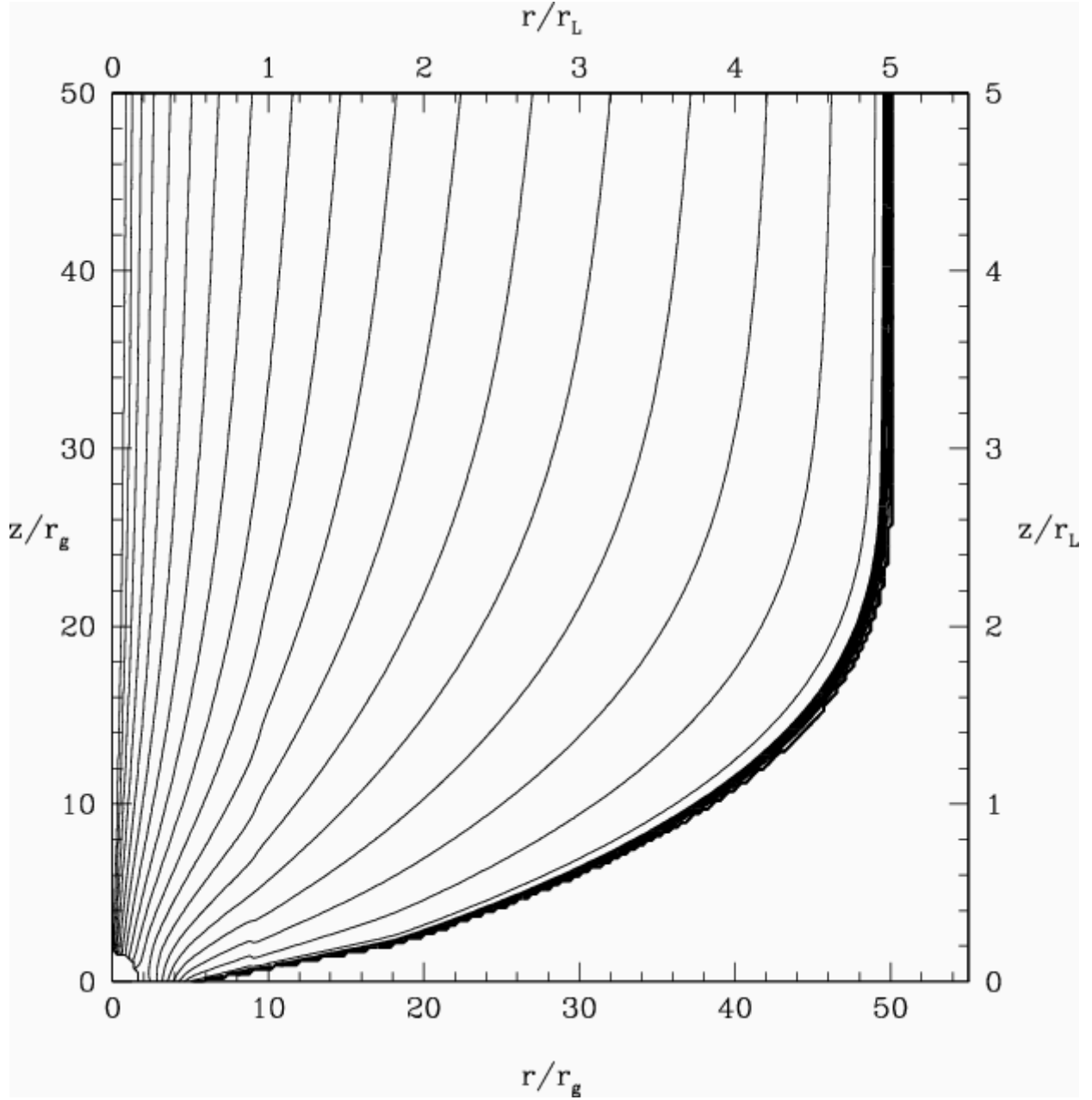}
\caption{The axisymmetric magnetic field structure of a jet from
a rotating black hole as solution of the Grad-Shafranov equation in
Kerr metric $\Psi(r,z)$.
The jet originates within $5\rg$ of the accretion disk.
Parameters:
BH angular momentum parameter $a=0.8$,
coupling constant $g_I = 0.717 $,
jet radius $r_{\rm jet} = 50\,r_{g}$,
asymptotic light cylinder $R_{\rm L} = 10\,r_{g}$,
black hole magnetic flux $\Psi_{\rm BH} = 0.1\,\Psi_{\rm total}$.
By prescribing $g_I$, $r_{\rm jet}$ and the disk flux distribution,
the shape of the collimating jet boundary ({\it thick curve})is determined
by the regularity condition at the light surface.
The inner boundary is the inner light surface around the ergosphere
(white hemisphere).
\label{fig_2D}
}
\end{figure}

\begin{figure*}

\epsscale{1.0}
\plotone{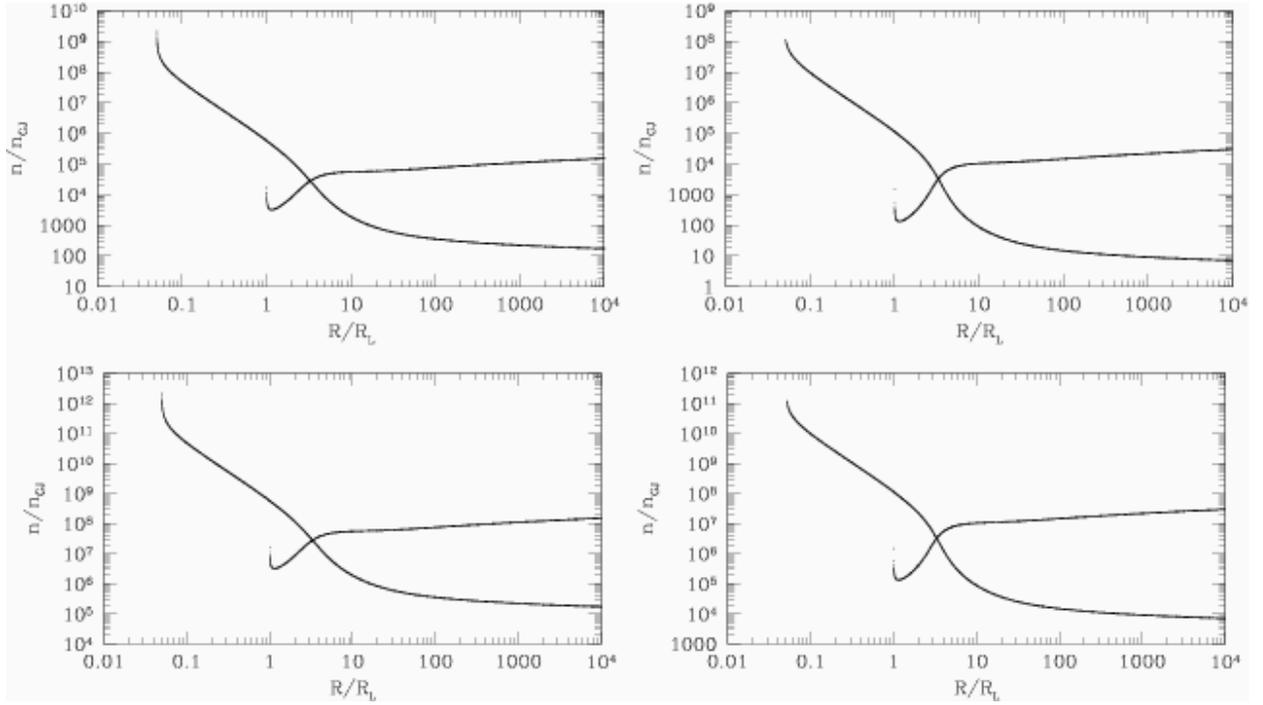}
\caption{Relativistic MHD wind/jet solution. Particle density in terms of
Goldreich Julian particle density as a function of radius along the magnetic
field line.
Parameter: $\sigm = 1000$ ({\it left}), $\sigm = 5000$ ({\it right}),
initial magnetic field strength $B_{z,\star} =  10^9{\rm G}$ ({\it top}),
$B_{z,\star} =  10^{12}{\rm G}$ ({\it bottom}).
\label{fig_n_ngj}
}
\end{figure*}

\begin{figure}

\epsscale{0.6}
\plotone{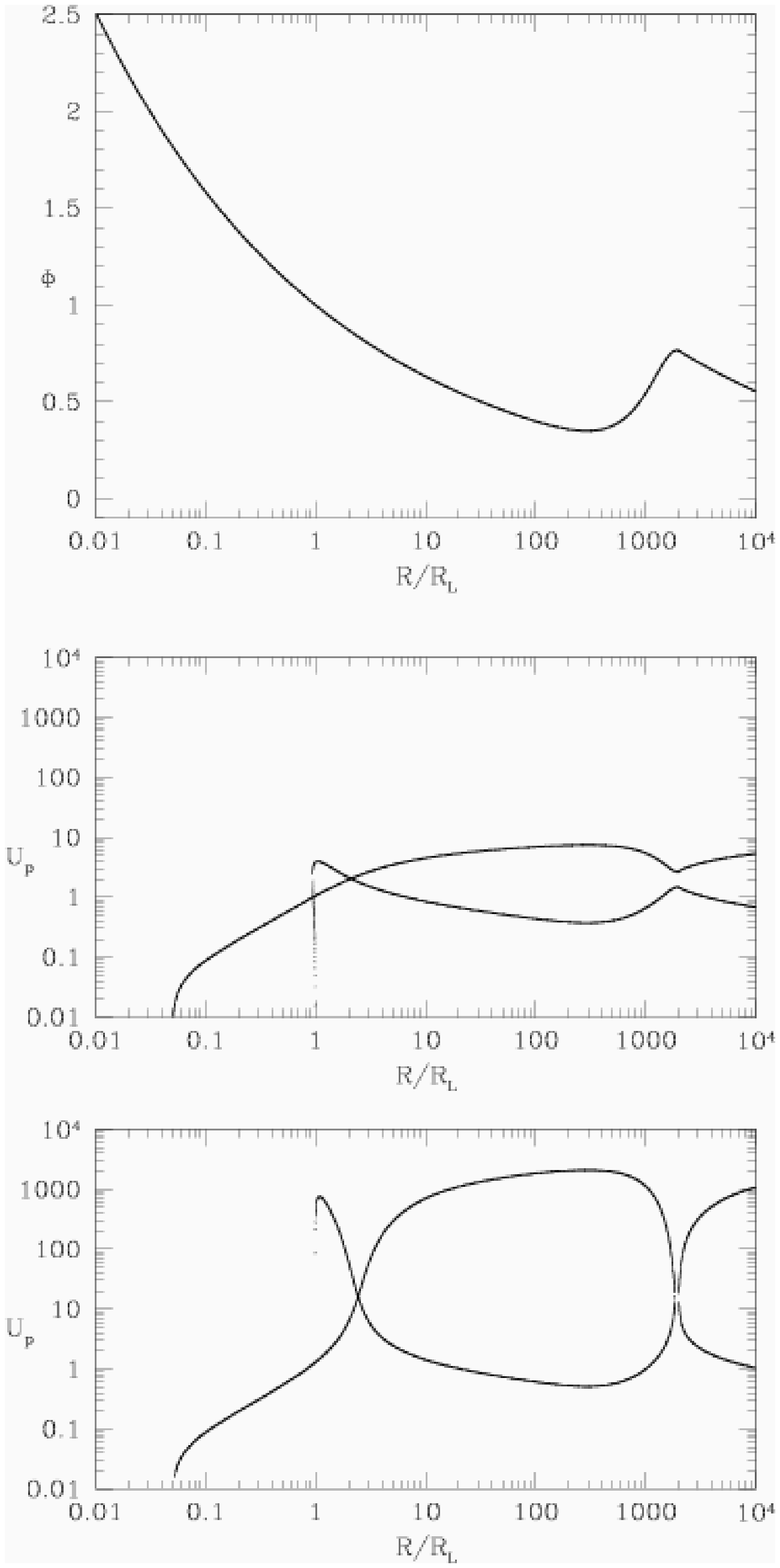}
\caption{Relativistic MHD wind/jet solution. Profiles of the magnetic flux
tube function ({\it top})
and the poloidal velocity ({\it below}) along the field line for a
re-collimating magnetic flux tube.
Magnetization $\sigma = 10, 5000$ (from {\it top} to {\it bottom}).
\label{fig_bottle}
}
\end{figure}

\clearpage 

\end{document}